\documentclass[proceedings]{JHEP3}
\usepackage{amsfonts}
\usepackage{amsmath}
\usepackage{epsfig,multicol}

\newbox\mybox

\newcommand\fverb{\setbox\mybox=\hbox\bgroup\verb}
\newcommand\fverbdo{\egroup\medskip\noindent\fbox{\unhbox\mybox}\ }
\newcommand\fverbit{\egroup\item[\fbox{\unhbox\mybox}]}
\conference{$\mathcal{PT}$-symmetric Deformations of the KdV Equation}
\abstract{We propose a new family of complex $\mathcal{PT}$-symmetric 
extensions of the Korteweg-de Vries equation. The deformed equations can be associated
to a sequence of non-Hermitian Hamiltonians. The first charges
related to the conservation of mass, momentum and energy are constructed. We
investigate solitary wave solutions of the equation of motion for 
various boundary conditions.}
\title{$\mathcal{PT}$-symmetric Deformations of the Korteweg-de Vries \\
Equation}
\author{Andreas Fring \\
Centre for Mathematical Science, City University\\
Northampton Square, London EC1V 0HB, UK\\
E-mail: \email{A.Fring@city.ac.uk}}

\input{tcilatex}

\begin{document}

\section{Introduction}

$\mathcal{PT}$-symmetry has served as a very fruitful guiding principle to
identify potentially interesting non-Hermitian Hamiltonians, which may
constitute physically relevant non-dissipative systems. The interest in
these type of configurations has started with a numerical observation made
in \cite{Bender:1998ke}, where it was found that the Hamiltonian 
\begin{equation}
H=p^{2}-g(iz)^{N+1}  \label{1}
\end{equation}
possess a real, positive and discrete eigenvalue spectrum for integers $%
N\geq 1$ with coupling constant $g\in \mathbb{R}^{+}$, despite it being
non-Hermitian $H\neq H^{\dagger }$ and unbounded from below, for $N=4n-1$
with $n\in \mathbb{N}$. The virtue of $\mathcal{PT}$-symmetry results from
the fact that whenever the Hamiltonian and the wavefunctions are left
invariant under a $\mathcal{PT}$-transformation the eigenvalues are
guaranteed to be real. However, the anti-linear nature of the $\mathcal{PT}$%
-operator is responsible for the fact that such a guarantee can not be
provided by the $\mathcal{PT}$-symmetry of the Hamiltonian alone \cite{EW,SW}%
. Unlike as for linear operators, for the $\mathcal{PT}$-operator its two
dimensional representation can be realized, in which case one speaks of
broken $\emph{PT}$-symmetry. One is then in a situation in which the
corresponding wavefunctions are not $\mathcal{PT}$-symmetric and the
eigenvalues occur in complex conjugate pairs. Nonetheless, even though $%
\mathcal{PT}$-symmetry of the Hamiltonian can not guarantee the reality of
the spectrum, it pre-selects a subclass of promising non-dissipative
systems. For recent results and a review see for instance \cite
{special,specialCzech,CArev}.

A couple of month ago Bender, Brody, Chen and Furlan \cite{BBCF} have
applied the above principle to identify interesting extensions of the
Korteweg-de Vries (KdV) equation \cite{KdV} 
\begin{equation}
u_{t}+uu_{x}+u_{xxx}=0.  \label{0KdV}
\end{equation}
The scaling properties of this equation for $x\rightarrow \alpha
x,t\rightarrow \beta t,u\rightarrow \gamma u$ are well known, e.g.~\cite
{Miura} 
\begin{equation}
\frac{\gamma }{\beta }u_{t}+\frac{\gamma ^{2}}{\alpha }uu_{x}+\frac{\gamma }{%
\alpha ^{3}}u_{xxx}=0,  \label{1kdv}
\end{equation}
and it has been remarked already at least thirty years ago that the KdV
equation remains invariant under a $\mathcal{PT}$-transformation $%
t\rightarrow -t,x\rightarrow -x$, see for instance p. 414 in \cite{Miura}.
This is of course just the particular case $\alpha =\beta =-\gamma =-1$.
However, this property has only been exploited in the above mentioned spirit
in \cite{BBCF}, where the KdV equation has been extended to the complex
domain in a $\mathcal{PT}$-symmetric manner 
\begin{equation}
u_{t}-iu(iu_{x})^{\varepsilon }+~u_{xxx}=0\ \ ~\ \ \ \ \text{\ ~~}%
\varepsilon \in \mathbb{R}.  \label{BKdV}
\end{equation}
One may think of equation (\ref{BKdV}) as being obtained from (\ref{0KdV})
by a scale invariant deformation 
\begin{equation}
u_{x}\rightarrow -\hat{q}(qu_{x})^{\varepsilon }\text{ \ \ \ \ \ \ \ \ }%
\varepsilon \in \mathbb{R},
\end{equation}
of the second term. When the deformation parameters scale as $q\rightarrow
\alpha /\gamma q$, $\hat{q}\rightarrow \gamma /\alpha \hat{q}$, equation (%
\ref{BKdV}) has the same behaviour under scaling as (\ref{1kdv}) for all
values of $\varepsilon $. The special case $q=\hat{q}=i$ yields a $\mathcal{%
PT}$-symmetric expression for $\alpha =\beta =-\gamma =-1$. Intriguingly,
the equation (\ref{BKdV}) were found to possess interesting solitary wave
solutions and two conserved charges were also constructed.

\section{A new $\mathcal{PT}$-symmetric deformation of the KdV equation}

It should be mentioned that complex extensions of the KdV equation have been
studied before, see e.g.~\cite{CKdV1,CKdV2,CKdV3} and in passing even some
special cases of equation (\ref{BKdV}) have been dealt with for instance in 
\cite{FSA}. However, only few properties have been studies for the latter
and $\mathcal{PT}$-symmetry has not been adopted as a guiding principle.
Motivated by the interesting findings in \cite{BBCF} and the usefulness of $%
\mathcal{PT}$-symmetric complex deformations in other contexts, see e.g.~
\cite{special,specialCzech,CArev}, we extend here its application. We
suggest that instead of deforming the second term in (\ref{1kdv}), by the
same principle one may equally well deform the last term or possibly all
terms. We shall demonstrate that the former case possesses some advantageous
features when compared with the previously outlined deformation.

Let us start by using the same $\mathcal{PT}$-symmetric deformation
principle 
\begin{equation}
u_{x}\rightarrow -i(iu_{x})^{\varepsilon }\text{ \ \ \ \ \ \ \ \ }%
\varepsilon \in \mathbb{R}  \label{pre}
\end{equation}
as employed in \cite{BBCF}, albeit now for the last term. This amounts to
replacing the third derivative as 
\begin{equation}
u_{xxx}\rightarrow i\varepsilon (iu_{x})^{\varepsilon -2}\left[ (\varepsilon
-1)u_{xx}^{2}+u_{x}u_{xxx}\right] .  \label{uxxx}
\end{equation}
In this way, simply applying (\ref{uxxx}) to (\ref{0KdV}), we obtain a new $%
\mathcal{PT}$-symmetric deformation of the KdV-equation 
\begin{equation}
u_{t}+uu_{x}+i\varepsilon (\varepsilon -1)(iu_{x})^{\varepsilon
-2}~u_{xx}^{2}+\varepsilon (iu_{x})^{\varepsilon -1}u_{xxx}=0.  \label{AF}
\end{equation}
At first sight the deformation (\ref{AF}) appears to be far less appealing
than the deformation (\ref{BKdV}). In the latter the effect of the
deformation was simply that the nonlinear term of the KdV-equation has
become somewhat more nonlinear, whereas in (\ref{AF}) we have replaced the
linear term by two highly nonlinear terms. Nonetheless, as a trade off the
deformation (\ref{AF}) has some very attractive features, which are not
present in (\ref{BKdV}). For instance, having a physical application in mind
we expect the deformed equation to be at least Galilean invariant just like
its undeformed counterpart (\ref{0KdV}). This property is lost in (\ref{BKdV}%
), but instead (\ref{AF}) is Galilean invariant, as it remains invariant
under the transformation 
\begin{equation}
x\rightarrow x-ct,\qquad t\rightarrow t,\qquad u\rightarrow u+c,  \label{Gal}
\end{equation}
where $c$ is the velocity of the moving reference frame. Furthermore, it was
difficult to construct conserved quantities for (\ref{BKdV}). Only two
charges could be constructed so far and in addition they turned out to be
complicated infinite series. We shall now demonstrate that this task is
surprisingly simple for (\ref{AF}), despite its high degree of nonlinearity
by relating it to a Hamiltonian formulation, which seems also impossible for
(\ref{BKdV}) as it appears to be a non-Hamiltonian dynamical system.

\section{$\mathcal{PT}$-symmetric deformations from a Hamiltonian formalism}

As we remarked, the $\mathcal{PT}$-symmetry analysis, which led to (\ref
{BKdV}), was carried out directly for the equation of motion. Recalling that
(\ref{1}) was obtained as a deformation of the standard harmonic oscillator
and that this principle has been applied to various Hamiltonian systems, it
appears highly desirable to perform deformations for the KdV system also on
the level of a Hamiltonian. This will enable us to relate these systems to
the arguments, which allow statements about the reality of the spectrum by
utilizing $\mathcal{PT}$-symmetry as outlined in the introduction. In the
equation of motion this property enters more indirectly and it is less clear
which kind of conclusions can be drawn from the symmetry property.

It is well known for a long time that the KdV-equation can be formulated as
a Hamiltonian system \cite{Gardner,Faddeev,Graham,Nutku}. Thus in this
spirit and in more direct analogy to the construction of (\ref{1}), we
propose to study the new non-Hermitian Hamiltonian density 
\begin{equation}
\mathcal{H=}u^{3}-\frac{1}{1+\varepsilon }(iu_{x})^{\varepsilon +1}~\ \ ~\ \
\ \ \text{\ }\varepsilon \in \mathbb{R}\text{.}  \label{Haf}
\end{equation}
For $\varepsilon \rightarrow 1$ we recover the standard Hamiltonian density
for the KdV-equation. Clearly $\mathcal{H}$ in (\ref{Haf}) is $\mathcal{PT}$%
-symmetric, since it remains invariant under the transformation: $%
t\rightarrow -t,x\rightarrow -x,i\rightarrow -i$ and $u\rightarrow u$.
Similarly as in the standard quantum mechanical setting, outlined in the
introduction, $\mathcal{PT}$-symmetry can be utilized to ensure the reality
of the energy $E$, which follows trivially with $\mathcal{H}(u(x))=\mathcal{H%
}^{\dagger }(u(-x))$%
\begin{equation}
E=\int\nolimits_{-a}^{a}\mathcal{H}(u(x))dx=-\int\nolimits_{a}^{-a}\mathcal{H%
}(u(-x))dx=\int\nolimits_{-a}^{a}\mathcal{H}^{\dagger }(u(x))dx=E^{\dagger }.
\end{equation}

Let us now derive the corresponding equation of motion by invoking the
variational principle for the Hamiltonian $H(u)=\int \mathcal{H}dx$ 
\begin{equation}
\frac{\partial u}{\partial t}=\frac{\partial }{\partial x}\left( \frac{%
\delta H(u)}{\delta u}\right) =\frac{\partial }{\partial x}\left( \frac{%
\delta \int \mathcal{H}dx}{\delta u}\right) =\frac{\partial }{\partial x}%
\left( \sum\nolimits_{n=0}^{\infty }(-1)^{n}\frac{d^{n}}{dx^{n}}\frac{%
\partial \mathcal{H}}{\partial u_{nx}}\right) .  \label{var}
\end{equation}
Evaluating (\ref{var}) for $\mathcal{H}$ in (\ref{Haf}) yields 
\begin{equation}
u_{t}+(-3u^{2}+\varepsilon (iu_{x})^{\varepsilon -1}~u_{xx})_{x}=0,
\label{KdVcon}
\end{equation}
or when not written as a conservation law 
\begin{equation}
u_{t}-6uu_{x}+i\varepsilon (\varepsilon -1)(iu_{x})^{\varepsilon
-2}~u_{xx}^{2}+\varepsilon (iu_{x})^{\varepsilon -1}u_{xxx}-\kappa =0,
\label{genKdV}
\end{equation}
where $\kappa $ is a constant. Note that (\ref{genKdV}) is almost (\ref{AF}%
), but corresponds to a deformation of the scaled KdV equation (\ref{1kdv}),
with $\alpha =\beta =1$, $\gamma =-6$ and $\kappa =0$, which, depending on
the context, is also frequently used in the literature for convenience.

\subsection{Integrals of motion and conserved quantities}

Having seen how to obtain the $\mathcal{PT}$-symmetrically deformed KdV
equation (\ref{AF}), or more precisely its scaled version (\ref{genKdV}),
from a Hamiltonian principle, we shall demonstrate next that is has further
interesting properties, which are absent in the deformation (\ref{BKdV}). As
mentioned, for (\ref{BKdV}) the authors of \cite{BBCF} could only construct
the two first conserved quantities in form of complicated infinite sums.
Here we find instead that for (\ref{genKdV}) these quantities can be
computed in a straightforward manner. Assuming to have a conserved quantity
of the form $\mathcal{I}^{(n)}=\int \mathcal{T}^{(n)}dx$, all we have to
verify is whether its Poisson bracket with the Hamiltonian is vanishing, see
e.g.~\cite{Gardner}. Viewing $\mathcal{I}^{(n)}(u)$ and $H(u)$ as
functionals of $u$ we have by definition 
\begin{equation}
\frac{d\mathcal{I}^{(n)}}{dt}=\int \frac{\delta \mathcal{T}^{(n)}}{\delta u}%
\frac{\partial u}{\partial t}dx=\int \frac{\delta \mathcal{T}^{(n)}}{\delta u%
}\left( \frac{\delta H}{\delta t}\right) _{x}dx=:\left\{ \mathcal{I}%
^{(n)},H\right\} .  \label{Poiss}
\end{equation}
Let us now employ (\ref{Poiss}) to establish that 
\begin{equation}
\mathcal{I}^{(1)}=\int udx,\qquad \mathcal{I}^{(2)}=\int u^{2}dx\qquad \text{%
and\qquad }\mathcal{I}^{(3)}=H(u),
\end{equation}
are indeed preserved under an evolution in time. We find that these
quantities are conserved when we invoke as standard boundary condition the
non-compact or compact case for $u,u_{x},\ldots $, that is being either
vanishing at infinity or periodic in space, respectively. This is easily
seen by computing 
\begin{eqnarray}
\frac{d\mathcal{I}^{(1)}}{dt} &=&\left\{ \mathcal{I}^{(1)},H\right\} =\int
(3u^{2}-\varepsilon (iu_{x})^{\varepsilon -1}~u_{xx})_{x}dx=0,  \label{e1} \\
\frac{d\mathcal{I}^{(2)}}{dt} &=&\left\{ \mathcal{I}^{(2)},H\right\} =\int
\left( 4u^{3}-\frac{2\varepsilon }{1+\varepsilon }(iu_{x})^{\varepsilon
+1}-2\varepsilon u(iu_{x})^{\varepsilon -1}u_{xx}\right) _{x}dx=0,
\label{e2} \\
\frac{d\mathcal{I}^{(3)}}{dt} &=&\left\{ \mathcal{I}^{(3)},H\right\}
=-\left\{ H,\mathcal{I}^{(3)}\right\} =0.  \label{e3}
\end{eqnarray}
The last conservation law follows trivially from the anti-symmetry property
of the Poisson brackets. We can also be more explicit and compute the
corresponding flux. Constructing vanishing Poisson bracket amounts to
seeking solutions of the conservation law 
\begin{equation}
\mathcal{T}_{t}^{(n)}+\mathcal{X}_{x}^{(n)}=0,
\end{equation}
with $-\mathcal{X}^{(n)}$ being the $n$th flux and $\mathcal{T}^{(n)}$ being
the $n$th conserved density. Then $\mathcal{I}^{(n)}=\int \mathcal{T}%
^{(n)}dx $ is a conserved charge provided the appropriate boundary
conditions hold. The case $n=1$ corresponds to the equation of motion itself
as can be read off directly from (\ref{KdVcon}). For the case $n=2$ we may
re-write (\ref{e2}) as a conservation law in the form 
\begin{equation}
\left( u^{2}\right) _{t}+\left( \frac{2\varepsilon }{1+\varepsilon }%
(iu_{x})^{\varepsilon +1}+2\varepsilon u(iu_{x})^{\varepsilon
-1}u_{xx}-4u^{3}\right) _{x}=0.
\end{equation}
We can also be more concrete about $\mathcal{T}^{(3)}=\mathcal{H}$ and
compute the associated flux 
\begin{equation}
\mathcal{X}^{(3)}=(\frac{\varepsilon ^{2}}{2}-\varepsilon
)(iu_{x})^{2\varepsilon -2}u_{xx}^{2}+3\left( \varepsilon
uu_{xx}-2u_{x}^{2}\right) u(iu_{x})^{\varepsilon -1}-i\varepsilon
(iu_{x})^{2\varepsilon -1}u_{xxx}-\frac{9}{2}u^{4},
\end{equation}
thus confirming (\ref{e3}). At this stage it is not clear whether there
exist higher conserved quantities. However, we suspect that similarly as for
most cases of the modified KdV equations and the generalized KdV equations
only three charges exist. We recall that the equation $%
u_{t}+u^{p}u_{x}+u_{qx}=0$ is only integrable, i.e.~possesses an infinite
amount of conserved quantities, for the cases $q=3,p=1,2$; $q=1,p\in \mathbb{%
N}$ and $q\in \mathbb{N},p=1$, see e.g.~\cite{Miura}.

\subsection{Solutions of the equations of motion}

We shall now construct solutions of the equations of motion (\ref{genKdV}).
One may expect to find a rich variety of different types of solutions
similarly as for the standard KdV equation. Over the years several methods
have been developed to find such solutions ranging from minimizing the sum
of the conserved charges \cite{Laxalmost}, the inverse scattering method 
\cite{Gardner:1967wc}, Hirota's bilinearization method \cite{Hirota}, etc.
Some methods demand as a prerequisite the model to be integrable. As this
feature is not guaranteed for the model at hand, in fact the conjecture is
that the model is not integrable, our aim is here just to obtain a first
impression in order to indicate that the above family of equations deserve
further attention. Following a simple procedure which has turned out to be
useful for the standard KdV equation, we may integrate (\ref{genKdV})
directly by assuming the solution to be a steady progressing wave 
\begin{equation}
u(x,t)=w(kx-\omega t)=v(x-ct),  \label{steady}
\end{equation}
with $c=\omega /k$. Substituting (\ref{steady}) into the equation of motion (%
\ref{genKdV}) yields after some straightforward manipulations 
\begin{equation}
v_{x}^{(n)}=e^{\frac{i\pi (4n+3\varepsilon +1)}{2(1+\varepsilon )}}\left[ 
\frac{\varepsilon +1}{\varepsilon }(v^{3}+\frac{c}{2}v^{2}+\kappa v+\hat{%
\kappa})\right] ^{\frac{1}{\varepsilon +1}},
\end{equation}
with $\hat{\kappa}$ being an additional constant of integration and $n$
labeling the various branches of the function. Separating variables then
yields 
\begin{equation}
x-ct=e^{\frac{i\pi (4n+\varepsilon -1)}{2(1+\varepsilon )}}\left( \frac{%
\varepsilon }{\varepsilon +1}\right) ^{\frac{1}{\varepsilon +1}}\int \frac{dv%
}{(v^{3}+\frac{c}{2}v^{2}+\kappa v+\hat{\kappa})^{1/(\varepsilon +1)}}.
\label{xt}
\end{equation}
Apart from computing the integral in (\ref{xt}), the main problem is here
that we need to solve the equation for $v$ in order to obtain $v(x-ct)$.
This is only possible in very few exceptional cases, but the knowledge of
the inverse function $(x-ct)(v)$ in some domain will be valuable as it
provides the information about the kind of general behaviour which is
possible. For convenience we choose the dispersion relation and the
constants of integration to be parameterized as 
\begin{equation}
c=4k^{2}(2-m),\quad \kappa =4k^{4}(1-m)\quad \text{and\quad }\hat{\kappa}=0.
\label{const}
\end{equation}
This choice is guided by the known solutions for $\varepsilon =1$ and leads
naturally to three qualitatively different cases.

\subsection{Analogues of the cnoidal solution}

Let us first recall how to solve equation (\ref{xt}) for the case $%
\varepsilon =1$, which should result into an elliptic integral as we
integrate the inverse of the square root of a cubic polynomial. With the
choice of constants (\ref{const}) we may bring (\ref{xt}) into the usual
form of an elliptic integral 
\begin{equation}
kx-\omega t=\pm \frac{k}{\sqrt{2}}\int\limits_{-2k^{2}}^{w}\frac{dt}{\sqrt{%
t^{3}+2k^{2}(2-m)t^{2}+4k^{4}(1-m)t}}=\pm \int\limits_{0}^{\phi (w)}\frac{%
d\theta }{\sqrt{1-m\sin ^{2}\theta }},  \label{ell}
\end{equation}
with $\phi (w)=\arcsin \sqrt{(1+w/2k^{2})/m}$. From (\ref{ell}) we deduce
therefore that $w(kx-ct)$ becomes the well known cnoidal solution for the
KdV equation 
\begin{equation}
u(x,t)=-2k^{2}\limfunc{dn}{}^{2}(kx-\omega t|m),  \label{cnoid}
\end{equation}
with $\limfunc{dn}$ being a Jacobian elliptic function depending on the
parameter $m\in \lbrack 0,1]$, see e.g.~\cite{Chan} for notation and
properties. As (\ref{cnoid}) indicates for $\varepsilon =1$, the cases $%
m=0,1 $ are special in general. For generic values of $\varepsilon $ we
evaluate (\ref{xt}) with the parameterization (\ref{const}) to 
\begin{eqnarray}
x-ct &=&e^{\frac{i\pi (4n+\varepsilon -1)}{2(1+\varepsilon )}}\left( \frac{%
v(1+\varepsilon )}{\varepsilon }\right) ^{\frac{\varepsilon }{1+\varepsilon }%
}\left( \frac{1}{4k^{2}(1-m)}\right) ^{\frac{1}{1+\varepsilon }}  \label{int}
\\
&&\times F_{1}\left( \frac{\varepsilon }{1+\varepsilon };\frac{1}{%
1+\varepsilon },\frac{1}{1+\varepsilon };\frac{1+2\varepsilon }{%
1+\varepsilon };\frac{-v}{2k^{2}(1-m)};\frac{-v}{2k^{2}}\right) . 
\notag
\end{eqnarray}
Here $F_{1}$ is the Appell hypergeometric function defined via a double
infinite sum as 
\begin{equation}
F_{1}(\alpha ;\beta ,\beta ^{\prime };\gamma
;x;y):=\sum\limits_{m=0}^{\infty }\sum\limits_{n=0}^{\infty }\frac{(\alpha
)_{n+m}(\beta )_{n}(\beta ^{\prime })_{m}}{n!m!(\gamma )_{n+m}}x^{n}y^{m}
\end{equation}
with $(\alpha )_{n}:=\prod\nolimits_{k=1}^{n}(\alpha +k-1)$. Since we can
not solve (\ref{int}) for $v$ let us plot $(x-ct)$ as a function of $v$ and
search for real solutions.

We depict our findings in figure 1. For $\varepsilon =1$ we recognize the
cnoidal solution (\ref{cnoid}). For clarity we did not indicate the
vanishing imaginary part in this case. For the other values of $\varepsilon $
we find always two different types of solutions. The first resembles
qualitatively the cnoidal solution and is either real for $v\in \lbrack
-1/2k^{2},0]$ or $v\in \lbrack 0,1/2k^{2}]$. In figure 1 we present $%
\varepsilon =3$; $n=2,4$; $k=1/\sqrt{2}$ for the former case and $%
\varepsilon =5$; $n=2,5$; $k=i/\sqrt{2}$ for the latter. The second type is
more similar to the tan$^{2}$ solution for $\varepsilon =1$ to be discussed
in the next section. These solution are real either for $v\in (-\infty ,0]$
or $v\in \lbrack 0,\infty )$. In figure 1 the former case is illustrated by $%
\varepsilon =3$; $n=2,4$; $k=i/\sqrt{2}$ and the latter by $\varepsilon =5$; 
$n=2,5$; $k=1/\sqrt{2}$.

\noindent \negthinspace \negthinspace \negthinspace \negthinspace
\negthinspace \negthinspace \negthinspace \negthinspace \negthinspace
\negthinspace \negthinspace \negthinspace \negthinspace \negthinspace
\negthinspace \negthinspace  \includegraphics[width=16.0cm]{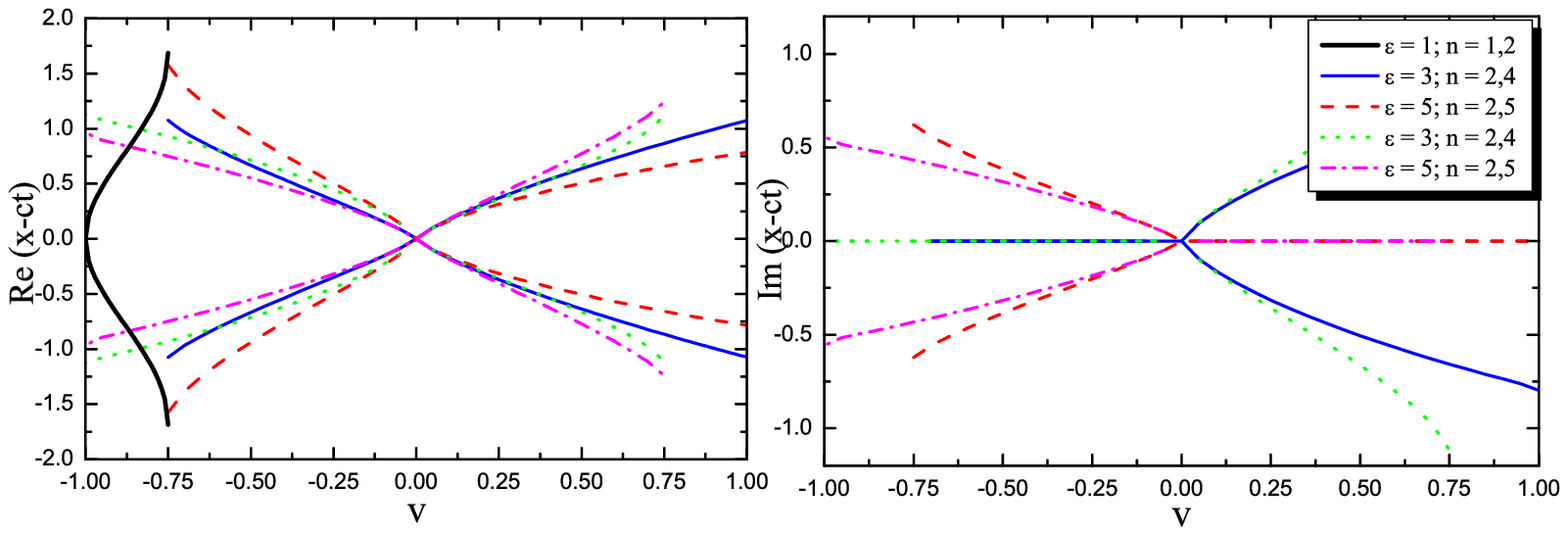}

\vspace*{-0.5cm}

\noindent {\small Figure 1: (x-ct) as a function of v for $\varepsilon
=1,3,5 $ for some particular branches.}

\subsection{Analogues of the tan$^{2}$ solution}

Next we consider the limit $m\rightarrow 0$. Keeping the parameterization (%
\ref{const}) we make the further convenient choice $k=\pm 1/\sqrt{2}$,
similarly as in the previous section, which amounts now to the boundary
condition $\kappa =1$. Then the Appell hypergeometric function $F_{1}$
reduces to the Gauss hypergeometric function $_{2}F_{1}$ defined as 
\begin{equation}
_{2}F_{1}(\alpha ;\beta ;\gamma ;x):=\sum\limits_{n=0}^{\infty }\frac{%
(\alpha )_{n}(\beta )_{n}}{n!(\gamma )_{n}}x^{n}=F_{1}(\alpha ;\beta
/2,\beta /2;\gamma ;x;x).
\end{equation}
Using furthermore the identity 
\begin{equation}
_{2}F_{1}(\alpha ;2\beta ;2\alpha +\beta ;x)=\alpha x^{-\alpha }B_{x}\left(
\alpha ,1-2\beta \right) ,
\end{equation}
where $B_{z}\left( \alpha ,\beta \right) $ is the incomplete beta function 
\begin{equation}
B_{z}\left( \alpha ,\beta \right) =\int\nolimits_{0}^{z}t^{\alpha
-1}(1-t)^{\beta -1},
\end{equation}
we obtain the simpler expression 
\begin{equation}
x-4t=e^{\frac{i\pi (4n+\varepsilon -1)}{2(1+\varepsilon )}}\left( \frac{%
\varepsilon }{\varepsilon +1}\right) ^{\frac{1}{\varepsilon +1}%
}{}B_{-v}\left( \frac{\varepsilon }{\varepsilon +1},\frac{\varepsilon -1}{%
\varepsilon +1}\right) .  \label{int2}
\end{equation}

For $\varepsilon =1$ (\ref{int2}) reduces further to $x-4t=\sqrt{2}\arctan
(\pm \sqrt{v})$, which may be solved for $v$, such that we obtain $%
u(x,t)=\tan ^{2}[(x-4t)/\sqrt{2}]$ as a solution for the standard KdV
equation. For generic values of $\varepsilon $ we depict (\ref{int2}) for
various values of the parameters in figure 2. For $\varepsilon =1$ we
perceive the real solution $x-4t=\sqrt{2}\arctan (\pm \sqrt{v})$ in panel
(a). A qualitatively similar type of solution is obtained for instance for
some branches for $\varepsilon =5$ as is seen also in panel (a). Panel (b)
confirms that for $v>0$ this solution is real. (The solid line is on top of
the dashed line) Interesting qualitatively different types of solutions are
obtained for instance for some branches for $\varepsilon =3,11$. We observe
from panel (c) that these solutions are very reminiscent of the one soliton
solution, to be discussed in the next section, albeit with the fundamental
difference that they are not vanishing asymptotically for large ($x-ct)$.
This is seen simply by using the property $B_{1}\left( \alpha ,\beta \right)
=\Gamma (\alpha )\Gamma (\beta )/\Gamma (\alpha +\beta )$ of the incomplete
beta function. For $v\rightarrow 1$ we obtain in (\ref{int2}) the definite
values $e^{\frac{i\pi (4n+\varepsilon -1)}{2(1+\varepsilon )}}\left( \frac{%
\varepsilon }{\varepsilon +1}\right) ^{1/(\varepsilon +1)}\Gamma (\frac{%
\varepsilon }{\varepsilon +1})\Gamma (\frac{\varepsilon -1}{\varepsilon +1}%
)/\Gamma (\frac{2\varepsilon -1}{\varepsilon +1})$. This limit is finite for
the parameter range except for $\varepsilon =1$, when $\lim\nolimits_{x%
\rightarrow 0}\Gamma (x)\rightarrow \infty $. In this case we obtain a
purely complex one soliton solution as can also be seen clearly in panel
(b). Having Galilean invariance for our equations, we may also move this
function as $v\rightarrow v+1$, such that the tails are located at $v=0$
rather than $v=-1$, which is a more familiar setting.

\noindent \negthinspace \negthinspace \negthinspace \negthinspace
\negthinspace \negthinspace \negthinspace \negthinspace \negthinspace
\negthinspace \negthinspace \negthinspace \negthinspace \negthinspace
\negthinspace \negthinspace  \includegraphics[width=16.0cm]{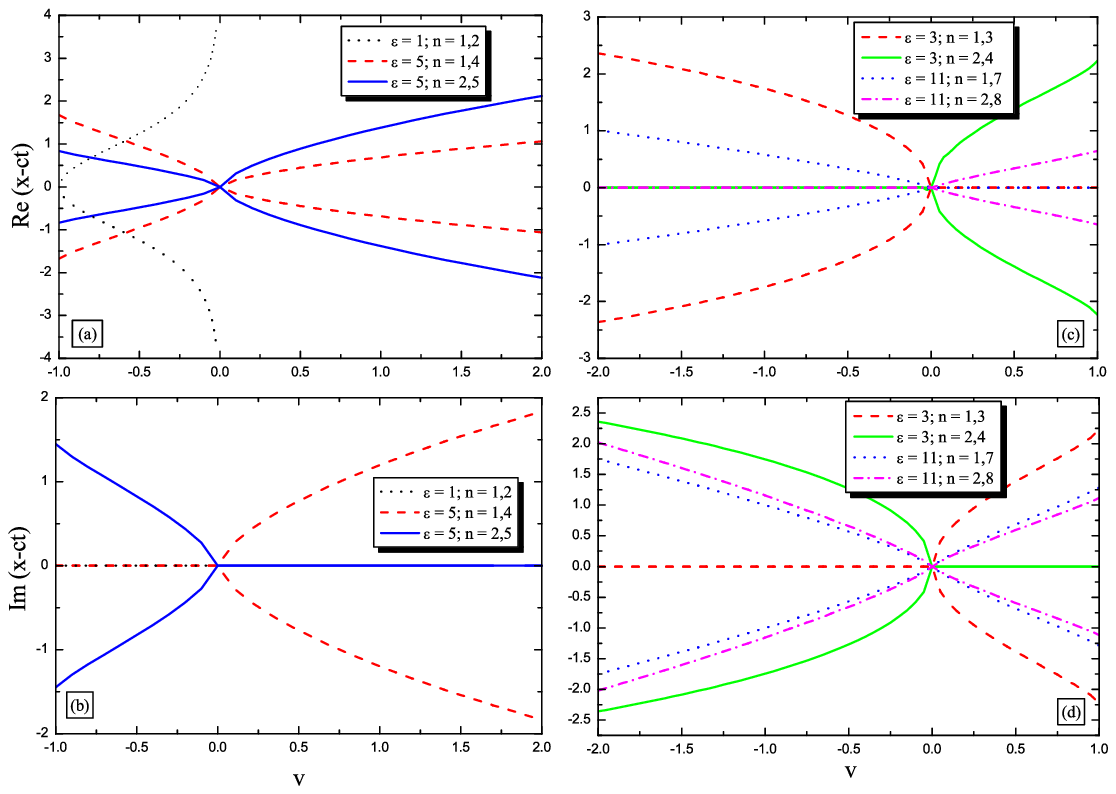}

\vspace*{-0.5cm}

\noindent {\small Figure 2: (x-ct) as a function of v for $%
\varepsilon =1,3,5$,$11$ for some particular branches and m=0.}

\subsection{Analogues of the one soliton solution}

Next we take the limit $m\rightarrow 1$ corresponding to the special case $%
\kappa =0$, which implements vanishing boundary conditions. Indeed, adopting
the parameterization (\ref{const}) the limit $m\rightarrow 1$ in (\ref{cnoid}%
) for $\varepsilon =1$ yields the asymptotically vanishing single soliton
solution $u(x,t)=-2k^{2}\func{sech}{}^{2}(kx-\omega t)$. Taking this limit
in (\ref{int}) for generic values of $\varepsilon $ gives 
\begin{equation}
x-ct=e^{\frac{i\pi (4n+\varepsilon -1)}{2(1+\varepsilon )}}\left( \frac{%
\varepsilon }{\varepsilon +1}\right) ^{\frac{1}{\varepsilon +1}}{}(2k^{2})^{%
\frac{1+\varepsilon }{2-\varepsilon }}~B_{-\frac{v}{2k^{2}}}\left( \frac{%
\varepsilon -1}{\varepsilon +1},\frac{\varepsilon }{\varepsilon +1}\right) .
\label{int3}
\end{equation}
We depict this function for various values of the parameters in figure 3.
The famous one soliton solution is clearly visible for $\varepsilon =1$. In
the other cases we obtain again two qualitatively different types of real
solutions. One type being real in the finite ranges $v\in \lbrack
-1/2k^{2},0]$ and $v\in \lbrack 0,1/2k^{2}]$ exemplified by $\varepsilon =5$%
; $n=1,4$; $k=1/\sqrt{2}$ and $\varepsilon =3$; $n=2,4$; $k=i/\sqrt{2}$,
respectively. The other type is real in the ranges for $v\in (-\infty ,0]$
and $v\in \lbrack 0,\infty )$, which we illustrated in figure 3 by $%
\varepsilon =3$; $n=1,3$; $k=i/\sqrt{2}$ and $\varepsilon =5$; $n=2,5$; $k=1/%
\sqrt{2}$, respectively.

\noindent \negthinspace \negthinspace \negthinspace \negthinspace
\negthinspace \negthinspace \negthinspace \negthinspace \negthinspace
\negthinspace \negthinspace \negthinspace \negthinspace \negthinspace
\negthinspace \negthinspace  \includegraphics[width=16cm]{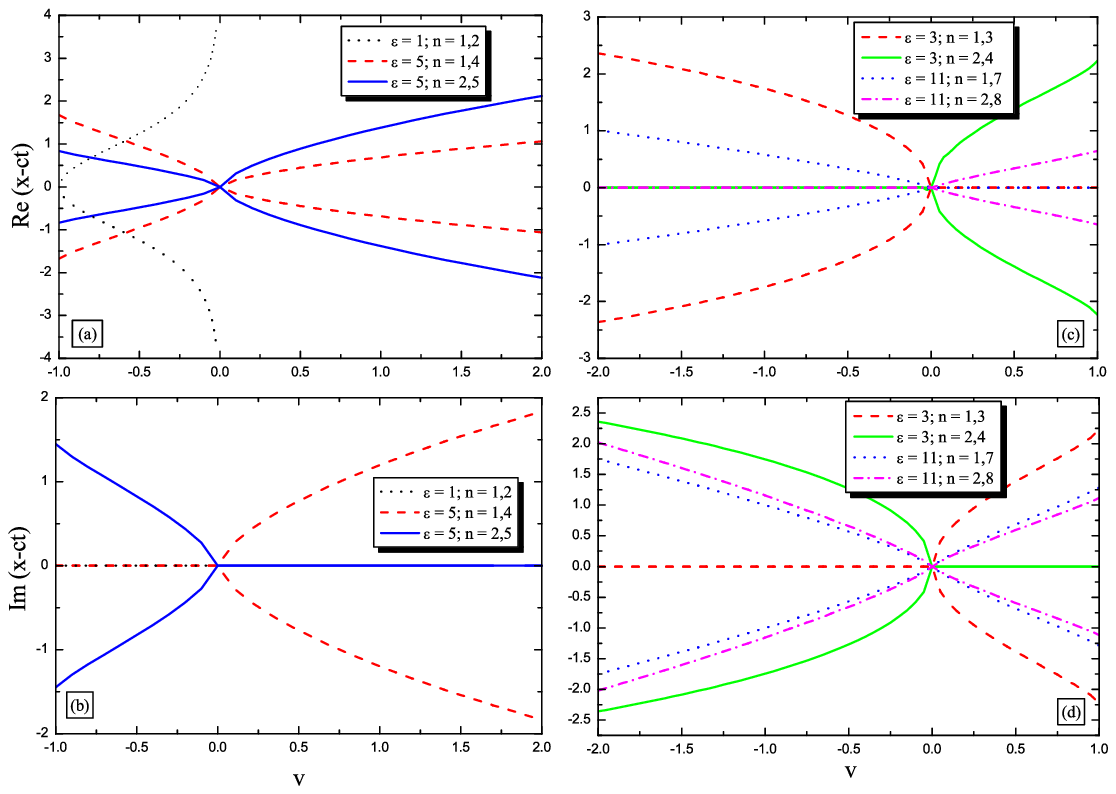}

\vspace*{-0.5cm}

\noindent {\small Figure 3: (x-ct) as a function of v for $\varepsilon =1,5$
and $\varepsilon =3,11$ with $k=\pm 1/\sqrt{2}$ and $k=\pm i/\sqrt{2}$,
respectively, for some particular branches with m=1.}

\section{Conclusions}

Alternatively to \cite{BBCF}, we proposed a new $\mathcal{PT}$-symmetric
complex deformed version of the KdV equation. The suggested deformation
allows for a simple non-Hermitian Hamiltonian formulation involving a
Hamiltonian density very reminiscent to the prototype $\mathcal{PT}$%
-symmetrically complex deformed quantum mechanical system (\ref{1}). The
model (\ref{Haf}) is Galilean invariant and three charges, related to the
conservation of mass, momentum and energy, together with their conservation
laws, were constructed. We demonstrated that there exist steady progressing
wave solutions for these models and identified analogues to the cnoidal and $%
\tan ^{2}$ solution. However, we did not find asymptotically vanishing
analogues to the one soliton solution.

Clearly there are many important questions left to be answered. It would be
interesting to establish that there exist three and only three charges for
the proposed deformation. Besides solving the equations more explicitly it
will be natural to seek for solutions on some rays in the complex plane. It
will be straightforward to extend these considerations to the modified KdV
and the generalized KdV equations. We shall leave these issues for future
investigations \cite{AMprep}.

\medskip

\noindent \textbf{Acknowledgments}. I am grateful to D.C. Brody for bringing
reference \cite{BBCF} to my attention. Discussions with C. Figueira de
Morisson Faria are gratefully acknowledged.

\end{document}